\documentclass[pra, superscriptaddress, twocolumn, showpacs, amsmath,amssymb, aps]{revtex4-1}

\usepackage{graphicx}
\usepackage{latexsym}
\usepackage{amsfonts}
\usepackage{hyperref}
\hypersetup{colorlinks=false, pdfborder={0 0 0}}
\usepackage[english]{babel}
\usepackage{bm}
\usepackage{units}
\usepackage{upgreek}
\usepackage{color}
\usepackage{todonotes}
\usepackage{tabularx}
\usepackage{soul}
\usepackage{lipsum}

\usepackage{braket}

\begin{document}

\title{Spectral analysis of two-dimensional Bose-Hubbard models}

\author{David Fischer}
\author{Darius Hoffmann}%
\affiliation{%
Institut f\"ur Theoretische Physik, Universit\"at Heidelberg, Philosophenweg 12, 69120 Heidelberg, Germany}%
\author{Sandro Wimberger}
\affiliation{Institut f\"ur Theoretische Physik, Universit\"at Heidelberg, Philosophenweg 12, 69120 Heidelberg, Germany}
\affiliation{Dipartimento di Fisica e Scienze della Terra, Universit\`a degli Studi di Parma, Via G. P. Usberti 7/a, 43124 Parma, Italy}
\affiliation{INFN, Sezione di Milano Bicocca, Gruppo Collegato di Parma, Italy}

\date{\today}

\begin{abstract}
One-dimensional Bose-Hubbard models are well known to obey a transition from regular to quantum-chaotic spectral statistics.
We are extending this concept to relatively simple two-dimensional many-body models. Also in two dimensions a transition from regular to chaotic spectral statistics is found and discussed. In particular, we analyze the dependence of the spectral properties on the bond number of the two-dimensional lattices and the applied boundary conditions. For maximal connectivity, the systems behave most regularly in agreement with the applicability of mean-field approaches in the limit of many nearest neighbor couplings at each site.
\end{abstract}

\pacs{03.65.Xp, 05.45.Mt, 71.35.Lk, 37.10.Jk}
\maketitle

\section{\label{sec:intro}Introduction}
\label{sec:intro}

State-of-the-art experimental techniques 
allow the experimentalist today to directly study a plethora of minimal models that have been developed in a solid-state physics context \cite{review}. Bose-Einstein condensates loaded into optical lattices, which realize spatially periodic structures, are used, e.g., to implement many-body models such as the celebrated Bose-Hubbard model (BHM) in one, two or even three dimensions. Other possible realizations of lattice systems with controllable couplings and interactions are based on arrays of optical cavities \cite{Oarray,2Doptic} or superconducting circuits \cite{Sarray}, for instance.

Already in 2003 and 2004, the spectral statistics of one-dimensional BHM were studied \cite{kol2003,kol2004}. Later extended models were also investigated, see e.g. \cite{Toma2007,BW2008,lubi2011,carlos2013,carlos2014}, which essentially confirmed these results. Recently, a semiclassical theory has been developed to understand the chaotic behavior of one-dimensional BHMs and to put the results mentioned above onto a firmer ground \cite{remy}. The knowledge of the spectral characteristics in a many-body system can be exploited to understand, engineer and control the system's dynamics, making such models and their experimental realizations interesting for the study of quantum thermalization and non-equilibrium transport phenomena, see e.g. \cite{kol2003,fine,carlos2014,therma1,therma2,schlagheck_therm}.

In this paper, we want to extend the spectral analysis to the case of two-dimensional BHMs with strong interparticle interactions. The dynamics of two-dimensional tight-binding systems was studied before in the noninteracting case \cite{kol2D-noINTER}, or in the specific case of a four mode interacting system \cite{2D-withINTER}. Analyzing different minimal models of up to a 3x3 square lattice, we will see how the geometry of the lattice and the number of permitted couplings (i.e. the number of bonds) determines the spectral properties of the systems. We restrict to such small numbers of lattice sites but treat particles numbers $N=6\ldots20$, such that the size of the total Hilbert space is sufficiently large for obtaining good statistics. This still allows for a full diagonalization of the many-body quantum problem and the scanning of a broad parameter space.

The next section presents our class of models and the studied lattices. It introduces all the spectral measures we computed as well. Section  \ref{sec:3} reports our main results, showing the transition from regular to chaotic spectral statistics in dependence on the lattices structures and the system parameters. Section \ref{sec:con} concludes the paper discussing also experimental ramifications of our work.

\section{Two-dimensional Bose-Hubbard-Models}
\label{sec:2}

The most general form of the types of Hamiltonians we are studying is
\begin{align}
	H= & \sum_{\bf R} \left( \epsilon_{\bf R} n_{\bf R}
	+ \tfrac{U}{2} n_{\bf R}(n_{\bf R}-1) \right) \nonumber \\
	- & \sum_{\bf R, R'} J_{\bf R, R'} (a^\dag_{\bf R} a_{\bf R'} + a^\dag_{\bf R'} a_{\bf R})
	\label{eq:bhm}
\end{align}

\begin{figure}[b]
\begin{center}
\includegraphics[width=1.05\linewidth]{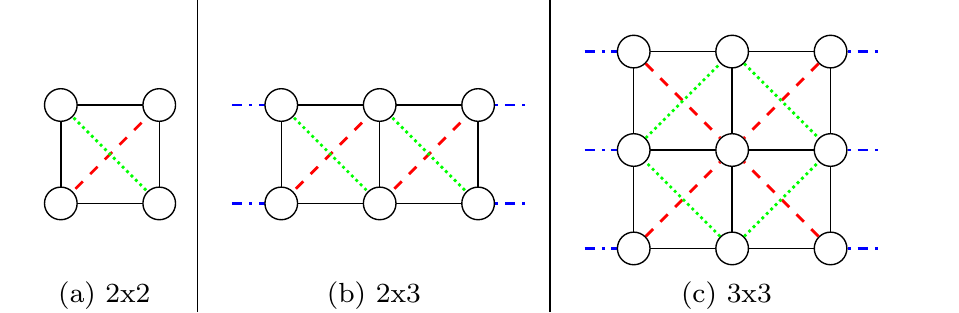}
\end{center}

\caption{\label{fig:1} Sketch of the lattice geometries used here. We study three basic settings of a 2x2 (a), a 2x3 (b) and a 3x3 lattice (c). The minimal bonds are always present (solid lines), while the additional diagonal bonds (green dotted and red dashed lines) are gradually added. In (b) and (c) we allow also for periodic boundary conditions (represented by the blue dot-dashed lines).}
\end{figure}

\begin{figure*}[t]
\begin{center}
\includegraphics[width=\linewidth]{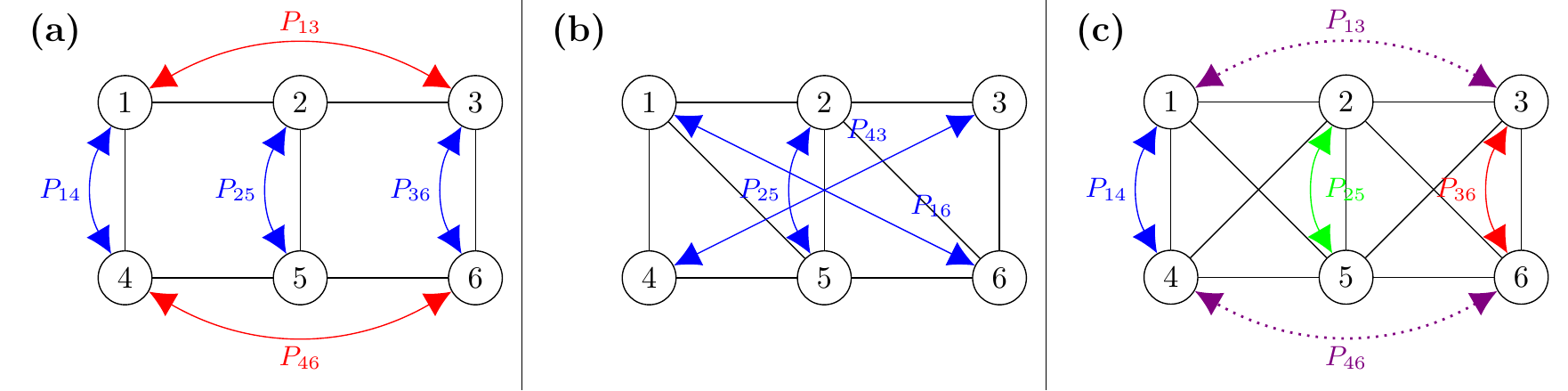}
\end{center}
\caption{\label{fig:2} Symmetries of our 2x3 lattice system - $P_{ij}$ denotes the transposition (exchange of the occupation numbers) of the corresponding sites: (a) without diagonals, two discrete symmetries: The exchange of the upper and lower row $P_{14} \cdot P_{25} \cdot P_{36}$ and the exchange $P_{13} \cdot P_{46}$. Since each of theses symmetries is self-inverse and therefore has the 2 eigenvalues $\pm 1$, the corresponding Hamiltonian decomposes in $2 \times 2$ subblocks. (b) with two diagonals, only a point-symmetry $P_{16} \cdot P_{25} \cdot P_{34}$ giving rise to two subblocks. (c) This case has most symmetries: each of the permutations $P_{14} $, $ P_{25} $, and $  P_{36}$ is a symmetry on its own, allowing a decomposition into $2 \times 2 \times 2$ blocks. There exists another symmetry $P_{13} \cdot P_{46}$, which commutes with $P_{25}$ but not with the other two. It therefore only decomposes the blocks corresponding to equal eigenvalues for $P_{14}$ and $P_{36}$. Since these are 4 of our 8 subblocks, we end up with 12 subblocks after the whole decomposition.
}
\end{figure*}
Here the vector ${\bf R} =(i,j)$ describes one point in the two-dimensional plane of a quadratic lattice, with the indices running from $i=1,\ldots,L$ and $j=1,\ldots,M$. The matrix of on-site energies $\epsilon_{\bf R}$ is assumed to be identical to zero. $a_{\bf R}$ and $a^\dag_{\bf R}$ are the annihilation and creation operators, respectively, and $n_{\bf R}=a^\dag_{\bf R}a_{\bf R}$ is the number operator. The specific geometries we investigate in the following are sketched in Fig. \ref{fig:1}. For all connected bonds the tunneling coupling $J_{\bf R, R'}=J=const.$, otherwise it is zero.

The particle number $N$ is chosen accordingly to the number of lattice sites $L\cdot M$ such that the total dimension $d$ of the many-particle Fock space is
\begin{equation}
	d= {N+L\cdot M-1\choose N} \,.
	\label{eq:fock-dim}
\end{equation}
Typical dimensions of the systems studied here are $d=1.8 \ldots 5.5 \times 10^3$, which provide us with sufficient statistics over the eigenvalues at numerically easily affordable computation times.

\subsection{\label{sec:level}Symmetries and unfolding}

When doing spectral statistics with Hubbard models, we must consider the symmetries of the systems, see e.g. \cite{kol2003,Toma2007,lubi2009}. Typical symmetries are translation invariance in the presence of period boundary conditions, mirror symmetries and parity. Which symmetries are simultaneously present depends on the form of the lattice and the number of bonds. In the case of the 2x3 lattice sketched in Fig. \ref{fig:1}(b), for instance, we have the symmetries discussed in detail in Fig. \ref{fig:2}. The four symmetry reduced subspaces shown in Fig. \ref{fig:2}(a) have dimensions ranging between $d_S=728 \ldots 784$, and in (b) we have $d_S=1491$ and $1512$ for the two subspaces respectively. In Fig. \ref{fig:2}(c) the dimensions of the subspaces vary between  $d_S=88 \ldots 434$. The larger variation of the dimensions in the latter case is due to the fact that some subspaces decompose further than others. The reason for this is the presence of an additional symmetry not commuting with all the other ones. Note that in the cases where diagonal bonds are present, the periodic boundary conditions also involve diagonal bonds such that translational invariance is respected.

With the right choice of the basis, the Hamiltonian of Eq. \eqref{eq:bhm} can be written in block-diagonal form corresponding to the irreducible representation that takes care of all symmetries which simultaneously commute with the Hamiltonian. Then we can focus our analysis on just one of these blocks so that level crossing due to states from different symmetry sectors are excluded. For large problems, this reduces the numerical complexity by about the number of independent blocks. For achieving optimal statistics we can analyse the individual blocks separately and then we collect in one plot all data thus obtained.

Before analyzing the eigenvalues of a symmetry reduced block, we must unfold the spectrum such that global trends in the spectrum are taken out. The effect of the unfolding procedure is that we can directly study the local spectral fluctuations which are crucial in any comparison with random matrix theory, see e.g. \cite{RMT} for details. The unfolding procedure we use is described as follows. After normalizing the spectral range to the interval $[0,1]$, we take the differences between consecutive energies $S_n=E_{n+1}-E_n$ and divide them by a local average: $s_n= S_n/\braket{S}_{W}$ where $\braket{\cdot}_W$ averages over the window $W=[n-l,n+l]$. The window size is typically $ l=5\ldots 25$ . We use in practice a smoothed energy level density, obtained by replacing the delta-functions in its definition by gaussian peaks of a width $\sigma$. Then the parameter $\sigma$ can be optimized such that we obtain good statistics without loss of relevant spectral information.

 
\subsection{\label{sec:res}Spectral measures}

After reducing the system into block-diagonal form, finding the eigenvalues by diagonalizing and unfolding the such obtained spectra, we analyze them using several methods which we are describing now. It is well known that symmetry-reduced (sub)systems can show regular, mixed and chaotic signatures which manifest in their spectral distributions \cite{RMT}. Typically, one compares them to the predictions coming from the theory of random matrices (RMT) \cite{RMT}. This is also what we do here. The standard measure for short-range spectral correlations is the distribution of the distances $s$ of nearest neighbor levels. For regular (pseudo-)random spectra, one expects a Poisson distribution
\begin{equation}
	P_{\rm Pois}(s)=e^{-s}\,.
	\label{eq:Pstat}
\end{equation}
Fully chaotic spectra do not allow for level crossings which are suppressed by level couplings. The corresponding distribution for our type of systems is the Wigner-Dyson one for time-invariant systems without spin (GOE):
\begin{equation}
	P_{\rm WD}(s)=\tfrac{s\pi}{2}e^{-\frac{\pi}{4}s^2}\,.
	\label{eq:WDstat}
\end{equation}
For mixed systems, an interpolating distribution is usually seen. One of the standard candidates is the so called Brody distribution with the Brody parameter $\beta$
\begin{equation}
	P_{\rm B}(s)=a(1+\beta)s^\beta e^{-as^{1+\beta}}\,,
	\label{eq:Brody}
\end{equation}
where $a\equiv \Gamma(\frac{2+\beta}{1+\beta})^{1+\beta}$.
The unfolding procedure mentioned in subsection \ref{sec:level} guarantees a mean level spacing $\braket{s}=1$. Hence, the numerical data is readily fitted with the normalized distributions above. We tested also the measure $P(r)$ introduced in \cite{rmass}. It is based on the ratio of consecutive nearest neighbor spacings $s_n$ and $s_{n-1}$, defined by
\begin{equation}
0< r_n = \frac{{\rm min}(s_n, s_{n-1}) }  { {\rm max}(s_{n},s_{n-1})} < 1\,.
\label{eq:rmass}
\end{equation}
This measure has the advantage that no unfolding is needed (because of the ratios of consecutive distances) and, therefore, statistical fluctuations are better controlled. The limiting case for Poisson spectra \cite{rmass} is
\begin{equation}
	P_{\rm Pois}(r)=\frac{2}{(1-r)^2}\,,
	\label{eq:Pr}
\end{equation}
and for GOE Wigner-Dyson spectra \cite{bogo}
\begin{equation}
	P_{\rm WD}(r) \sim \frac{27}{4}\frac{r+r^2}{(1+r+r^2)^{5/2}}\,.
	\label{eq:PWD}
\end{equation}

To test also the long-range correlations in the spectra we analyze in parallel to the short-range measures mentioned above the number variance of levels defined by
\begin{equation}
	\Sigma^2(L)= \bigg \langle  \left(  N(E,E+L) - L )\right)^2 \bigg \rangle_{E}\,,
	\label{eq:sigma}
\end{equation}

\begin{figure}[b]
\centering
\includegraphics[width=1.1\linewidth]{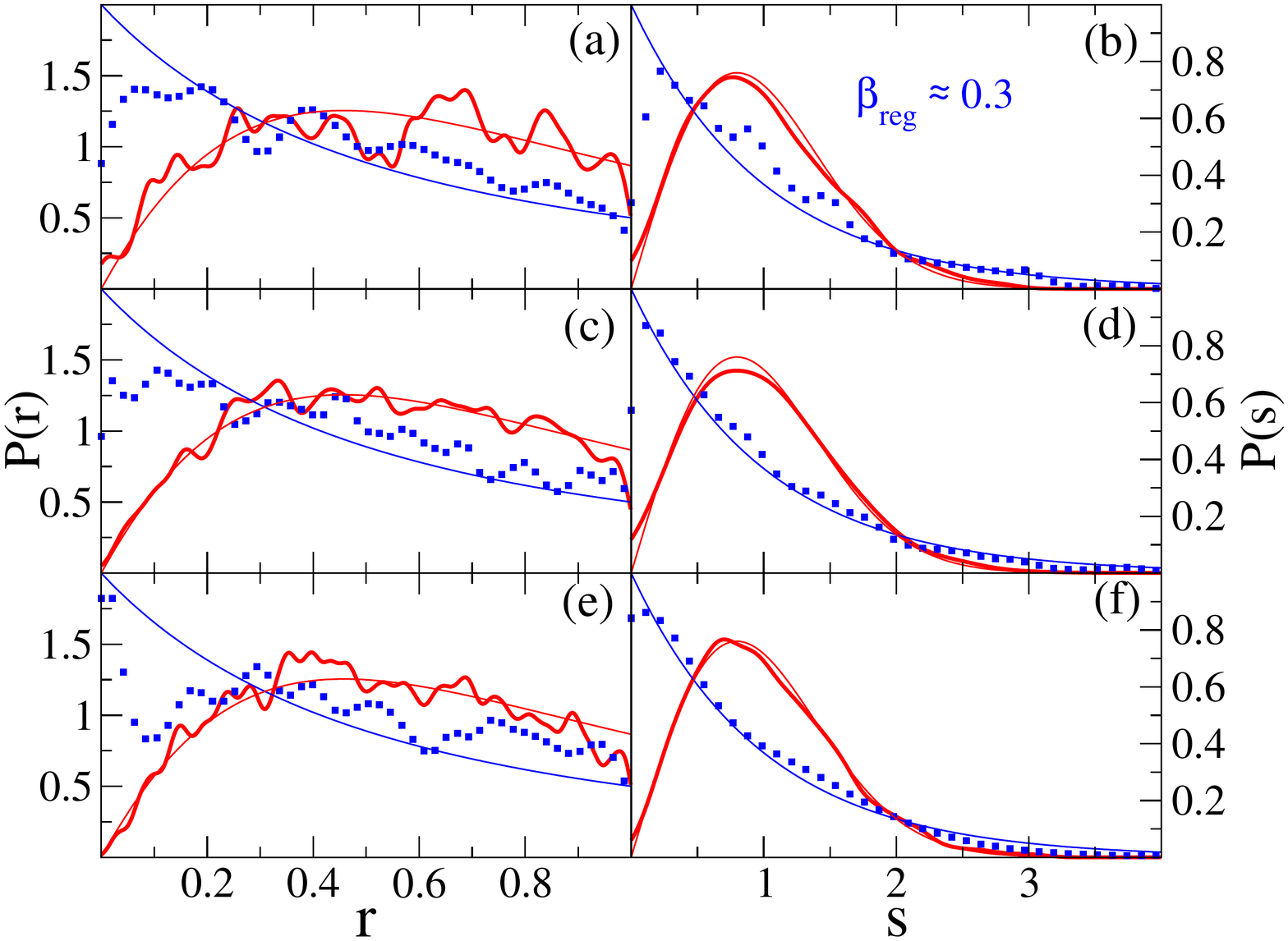}
\caption{Spectral analysis of 2D BHMs using the measures $P(r)$ (left panels: (a), (c), (e)) and $P(s)$ (right panels: (b), (d), (f)). Each panel exemplifies one regular (blue/black dotted lines) and one chaotic case (red/gray solid lines).The lattices structures are 2x2  (upper panels, $N=20$ with $U/J=2.42$ for regular and $U/J=0.87$ for chaotic case), 2x3 (middle panels, $N=10$, $U/J=15.0$ and $U/J=1.51$), and 3x3 (lower panels, $N=6$, $U/J=1000$ and $U/J=1.71$), all with open (Dirichlet) boundary conditions and without diagonal couplings. The RMT predictions are shown by the dashed lines by comparison, c.f. Eq. \eqref{eq:Pstat}-\eqref{eq:WDstat} (left) and \eqref{eq:Pr}-\eqref{eq:PWD} (right). The transition from regular to chaotic statistics is clearly visible. In the shown case it is controlled by the system parameter $U/J$.
In panel (b), for the regular case, we do not have a fully regular spectrum for this parameter $U/J$. This fact is indicated by a Brody parameter $\beta \approx 0.3$ significantly larger than zero, c.f. Eq.  \eqref{eq:Brody}.
}
\label{fig:3}
\end{figure}
where $\braket{\cdot}_{E}$ denotes the average over the studied energy interval. $N(E,E+L)$ is the number of unfolded levels with the interval $[E,E+L]$.
RMT predicts \cite{RMT}
\begin{equation}
	\Sigma^2_{\rm Pois}(L)=L 
	\label{eq:sigma_reg}
\end{equation}
and
\begin{equation}
	\Sigma^2_{\rm WD}(L)=\frac{2}{\pi ^2}\left(\ln(2\pi L)+\gamma + 1 - \frac{\pi ^2}{8}\right) + \mathcal{O}(\frac{1}{L})\,,
	\label{eq:sigma_chaos}
\end{equation}
respectively. $\gamma \approx 0.57722$ is the Euler constant. 

Finally, in order to scan over a wide range of parameters, it turned out that it is useful to quantify the spectral correlations by a single number (i.e. averaged over an entire spectrum). We use here a $\chi^2$ test to measure the distance from the obtained numerical data to the theoretical predictions from the 
Eqs. \eqref{eq:Pstat} and \eqref{eq:WDstat}, respectively:
\begin{equation}
\chi^2_{*} \equiv \log_{10}\bigg\{\int_0^\infty ds \quad \bigg(P_{*}(s)-P_{\text{numerical}}(s)\bigg)^2 \bigg\}.
\end{equation}
Here * stands for Poisson (Pois) or Wigner-Dyson (WD) respectively.
 Moreover, we define the new measure based on the number variance:
\begin{equation}
	L_{\rm rel}(q) \equiv  \max\{L:\frac{|\Sigma^2(L)-\Sigma^2_{\rm WD}(L)|}{\Sigma^2_{\rm WD}(L)} < q\} \, .
	\label{eq:CL}
\end{equation}
This correlation length characterizes up to which $L$ the numerical $\Sigma^2(L)$ follows the predictions for quantum chaos. Since the Poissonian case corresponds to a totally uncorrelated spectrum, we can indeed interpret it as a measure of the correlation length in the spectrum. The 'typical' choice $q=1$ is justified by the fact that due to the unfolding procedure the energy scale of distances is normalized to unity.

\section{Spectral analysis}
\label{sec:3}

\begin{figure}[b]
\includegraphics[width=1\linewidth]{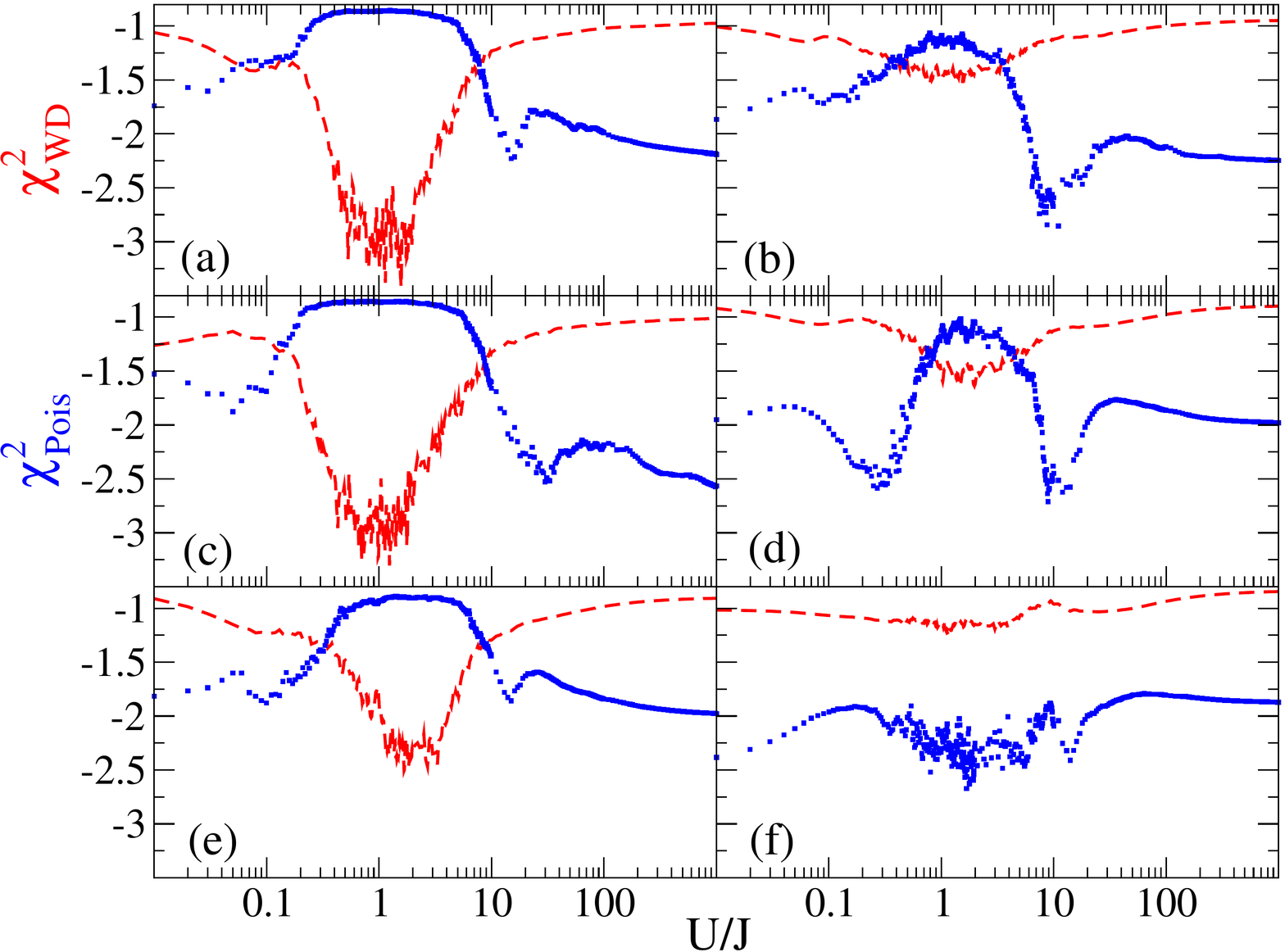}
\caption{$\chi^2$ statistical tests for the regular-to-chaotic transition for a 2x3 lattice at $N=10$. Small $\chi^2_{\rm Pois}$ and $\chi^2_{\rm WD}$ values indicate a good correspondence with $P_{\rm Pois}(s)$ or $P_{\rm Pois}(s)$, respectively. (a,b) show results for no diagonal coupling in Fig. \ref{fig:1}, (c,d) for the red dashed diagonal couplings, and (e,f) for a maximal number of diagonal couplings at open (a,c,e) and periodic (b,d,f) boundary conditions.}
\label{fig:4}
\end{figure}

A first overview over the spectral behavior of the systems is presented in Fig. \ref{fig:3}. Only open (Dirichlet) boundary conditions are applied in theses cases. Please note that the Hamiltonian \eqref{eq:bhm} effectively has just one parameter since we may divide by $J$ for instance. Hence we denote the control parameter here by $U/J$. The transition from regular to chaotic statistics is clearly visible and controlled by $U/J$ at fixed particle number $N$ and at fixed lattice structure. Because of the unfolding, the definition of the parameter $r$, see Eq. \eqref{eq:rmass}, and since $U/J$ is dimensionless all plotted quantities are dimensionless in the following.

For the next plot we pick the structure of the 2x3 lattice as a paradigm for the other geometries. Figure \ref{fig:4} shows the relative deviations of the numerically obtained distributions $P(s)$ from the regular and chaotic expectations, respectively. The data is for fixed particle number $N=10$, for a large window of $U/J$ and various configurations of bonds and different boundary conditions. We observe that the spectral characteristics are controlled by the number of bonds to the nearest neighbors.
Figure \ref{fig:4} highlights, in particular, that the systems are much more regular when periodic boundary conditions are applied, i.e. when more bonds are effectively present. Moreover, the panels (a-c) in Fig. \ref{fig:4} confirm the expectation of most chaoticity in the region around $U/J \approx 1$ at average atom numbers per site of order one \cite{kol2003,kol2004,Toma2007,BW2008,lubi2011,carlos2013,carlos2014}.

\begin{figure}[b]
\includegraphics[width=1.05\linewidth]{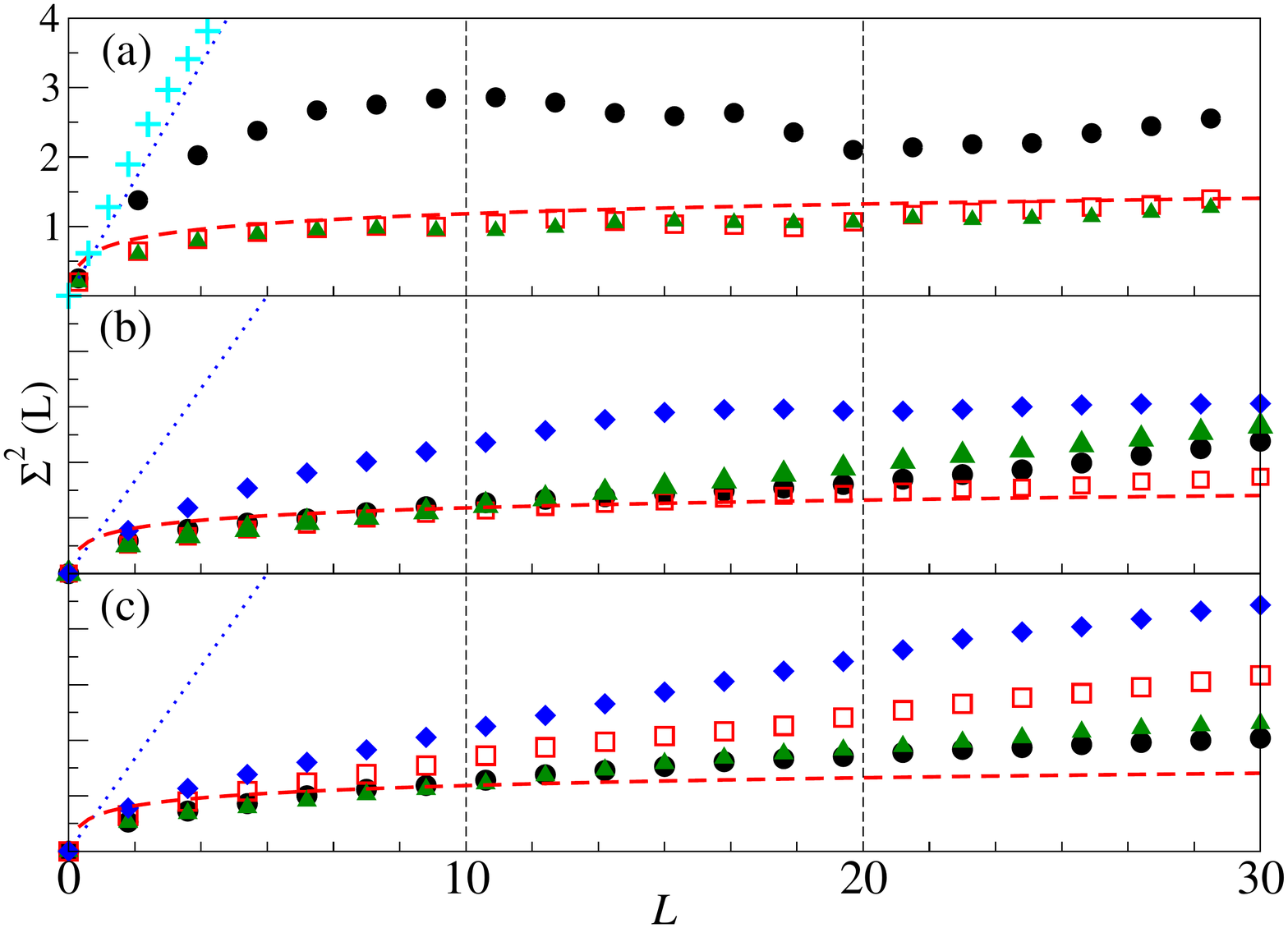}
\caption{$\Sigma^2$ measure for the three cases: (a) 2x2 lattice $N=20$; (b) 2x3 lattice $N=10$; (c) 3x3 lattice $N=6$.
The control parameter $U/J$ is chosen to minimize $\chi^2_{WD}$. Displayed in each panel are data for: 
lattice without diagonal coupling (triangles) with $U/J=0.87$ (a), $U/J=1.57$ (b) and $U/J=1.71$ (c) --
with one diagonal per plaquette, i.e. per square unit cell, (squares) at $U/J=0.955$ (a), $U/J=0.88$ and (c) $U/J=1.92$ --
maximal number of bonds at open boundary conditions (filled circles) at $U/J=1.385$ (a), $U/J=1.67$ (b) and $U/J=1.65$ (c) --
and periodic boundary conditions without diagonals (filled diamonds) at $U/J=0.79$ (b) and $U/J=3.3$ (c).
In (a) also a regular case with no diagonal bonds and $U/J=6.136$ (crosses) is presented for comparison.
%
%
%
%
 }
\label{fig:5}
\end{figure}

After the short-range correlations, we analyze also the long-range correlations. Figure \ref{fig:5} collects numerical results in particular for quantum chaotic spectra. The various plots show data for the three lattices sketched in Fig. \ref{fig:1}, again for different numbers of nearest-neighbor bonds. 
Here the systems with connectivity in Fig. \ref{fig:5} (a) and in addition with periodic boundary conditions in (b) and (c) are also worst approximated by the chaotic prediction from RMT.

\begin{figure*}[t]
\includegraphics[width=0.9\textwidth]{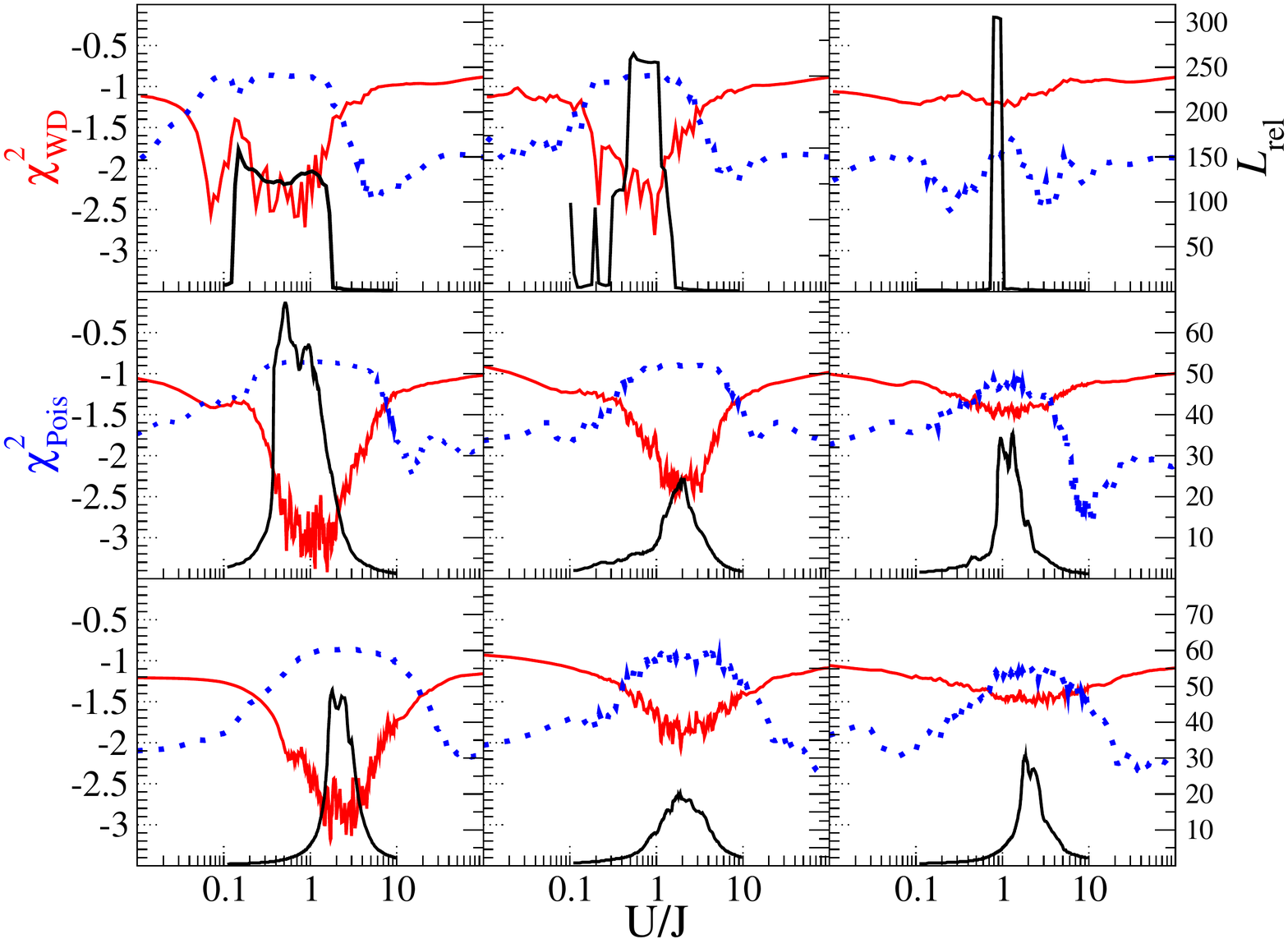}
\caption{Overview over the $\chi^2$ tests (left axes) and the analysis of long-range correlations with $L_{\rm rel}(q=1)$ from Eq. \eqref{eq:CL} (right axes). Lattices structures: 2x2 with $N=20$ (upper panels), 2x3 with $N=10$ (middle panels), and 3x3 with $N=6$ (lower panels). The left panels are for the minimal couplings without any diagonal bonds in Fig. \ref{fig:1}. The central column shows the cases of one diagonal bond (upper) and four diagonals with open boundary conditions (middle and lower panel). For the lower middle panel this corresponds to the red dashed lines in Fig \ref{fig:1}(c). The right panels are for two diagonals in the case 2x2 (upper), and no diagonals with periodic boundary conditions (middle and lower panel).}
\label{fig:6}
\end{figure*}

Our systems--represented by the Hamiltonian of \eqref{eq:bhm} and the geometries sketched in Fig. \ref{fig:1}--allow for a large variety of possible choices of parameters ($N$ and $U/J$) and bond numbers (including the choice of boundary conditions). Fig. \ref{fig:6} collects most of the possible cases for fixed $N$ at a given lattice structure. The individual plots show scans over a wide range of $U/J$. The different lines present the $\chi^2$ tests for the correspondence with the chaotic $P_{\rm WD}(s)$ (blue dotted) and the regular $P_{\rm Pois}(s)$ (red dashed). Also our newly introduced measure $L_{\rm rel}(q=1)$ is shown by the black solid lines. In all cases, we find a more or less broad chaotic region, for which $UN/J \sim 1$, a well known condition for strong mode coupling in the one-dimensional BHM \cite{kol2003,kol2004,Toma2007,BW2008,lubi2011,carlos2013,carlos2014,remy}. Both measures $\chi^2$ and $L(q)$ agree in indicating the most chaotic regions; and again the correspondence is optimal for the cases shown in the leftmost panels in Fig. \ref{fig:6}. This nicely corroborates the results of Fig. \ref{fig:5}. The chaotic regions are the less broad the more bonds are present in the structures. This fact is in agreement with the well-known expectation that mean-field approaches typically work best when all sites are coupled to as many nearest neighbors as possible, see, e.g., Ref. \cite{franz} in the context of generalized one-dimensional BHMs.

\section{\label{sec:con}Discussion and conclusions}

In this paper we analyzed the spectra of a class of small two-dimensional BHMs, scanning a broad range of parameters and lattices structures.
We found a transition from regular-to-chaotic behavior in almost all cases which is controlled by the system parameters $U/J$ and $N$ as well as by the number of coupled bonds in the models. Our work naturally extends studies of the one-dimensional BHM \cite{kol2003,kol2004,Toma2007,BW2008,lubi2011,carlos2013,carlos2014,remy,schlagheck_therm} and related Hubbard models for fermonic particles \cite{1-2DFermi}.

Minimal systems as investigated here could be realized experimentally by coupled one-dimensional chains of BHMs, see, e.g., Ref. \cite{2Dchains} for recent experiments in this direction, or by optical two-dimensional lattices, see a similar theoretical proposal in Ref. \cite{2Doptic}. Such experimental implementations could study quantum chaos of simple but strongly interacting indistinguishable bosons, complementary to recent scattering experiments with more complex ultracold erbium atoms \cite{QC-cold}.

\begin{acknowledgments}
SW acknowledges financial support by the FIL program of Parma University. We thank R\'emy Dubertrand for valuable discussions.
\end{acknowledgments}

\end{document}